\documentclass[aip,graphicx]{revtex4-1}
\usepackage{graphicx}

\begin{document}


\title{Magnified Time-Domain Ghost Imaging} 



\author{Piotr Ryczkowski}
\email[]{piotr.ryczkowski@tut.fi}
\affiliation{Optics Laboratory, Tampere University of Technology, Tampere, Finland}

\author{Margaux Barbier}
\affiliation{Optics Laboratory, Tampere University of Technology, Tampere, Finland}
\affiliation{current affiliation: FOTON Laboratory, CNRS, University of Rennes 1, ENSSAT, 6 rue de Kerampont, F-22305 Lannion, France}

\author{Ari T. Friberg}
\affiliation{Department of Physics and Mathematics, University of Eastern Finland, Joensuu, Finland}

\author{John M. Dudley}
\affiliation{Institut FEMTO-ST, UMR 6174 CNRS-Universit\'e de Franche-Comt\'e, Besan\c con, France}

\author{Go\"ery Genty}
\affiliation{Optics Laboratory, Tampere University of Technology, Tampere, Finland}

\date{\today}

\begin{abstract}
Ghost imaging allows to image an object without directly seeing this object. Originally demonstrated in the spatial domain using classical or entangled-photon sources, it was recently shown that ghost imaging can be transposed into the time domain to detect ultrafast signals with high temporal resolution. Here, using an incoherent supercontinuum light source whose spectral fluctuations are imaged using spectrum-to-time transformation in a dispersive fiber, we experimentally demonstrate magnified ghost imaging in the time domain. Our approach is scalable and allows to overcome the resolution limitation of time-domain ghost imaging.
\end{abstract}

\pacs{}

\maketitle 

\section{Introduction}
Ghost imaging allows the indirect retrieval of the image of an object illuminated by a spatially-structured pattern. The image is obtained from the correlation between the spatially-resolved structured illumination pattern and the total intensity transmitted through (or reflected by) the object\cite{Erkmen10,Bennink04}. Ghost imaging has been extensively studied in the spatial domain since the mid-1990s, using various types of light sources ranging from spatially-entangled photons sources\cite{Bennink04,Klyshko88_1,Klyshko88_2,Pittman95,Abouraddy01} to spatially incoherent classical light sources\cite{Bennink04,Bennink02,Scarcelli06,Meyers08,Shirai11,Zhang14} and, more recently, pre-programmed illumination by a spatial light modulator\cite{Sun13}. More advanced schemes based on multiplexing have also been demonstrated to reduce the acquisition time\cite{Zhang2015} or to image objects which vary slowly with time\cite{Devaux2016}. Compared to standard imaging techniques, a unique property of ghost imaging is its insensitivity to distortions that may occur between the object and the single-pixel detector that only measures the total transmitted (or reflected) intensity\cite{Ferri05,Meyers11}. This inherent insensitivity to external perturbations makes ghost imaging particularly appealing for long range applications such as e.g. LIDAR or atmospheric sensing.
Recently, exploiting space-time duality in optics\cite{Tournois64,Kolner89,Kolner94,Salem13}, ghost imaging was transposed into the time domain to produce the image of an ultrafast signal by correlating in time the intensity of two light beams, neither of which independently carried information about the signal\cite{Ryczkowski16}. Significantly, it was also demonstrated that the technique is insensitive to distortion that the signal may experience between the object and the detector e.g. due to dispersion, nonlinearity, or attenuation. A potential important limitation of ghost imaging in the time-domain is the finite resolution determined by the fluctuation time of the random light source and/or the speed of the detection system that measures the random intensity fluctuations. Here, we improve significantly the resolution of ghost imaging in the time-domain by reporting a new proof-of-concept experimental setup that allows to generate a magnified ghost image of an ultrafast waveform.
Our approach is inspired by shadow imaging in the spatial domain and builds on the dispersive Fourier transform of the fast fluctuations of an incoherent supercontinuum (SC). Dispersive Fourier transform uses the group-velocity dispersion of optical fibers to convert into the time domain spectral fluctuations\cite{Goda13}, and it has been successfully applied in the past to single-shot spectral studies of nonlinear instabilities in fiber optics\cite{Solli12}, or to perform analog-to-digital conversion and dynamic imaging\cite{Goda13}. By improving significantly the resolution of time-domain ghost imaging, our results open a new avenue to blindly detect and magnified ultrafast signals. 

\section{Magnified time-domain ghost imaging using Dispersive Fourier Transform}
\begin{figure}
	\includegraphics[width=.7\linewidth]{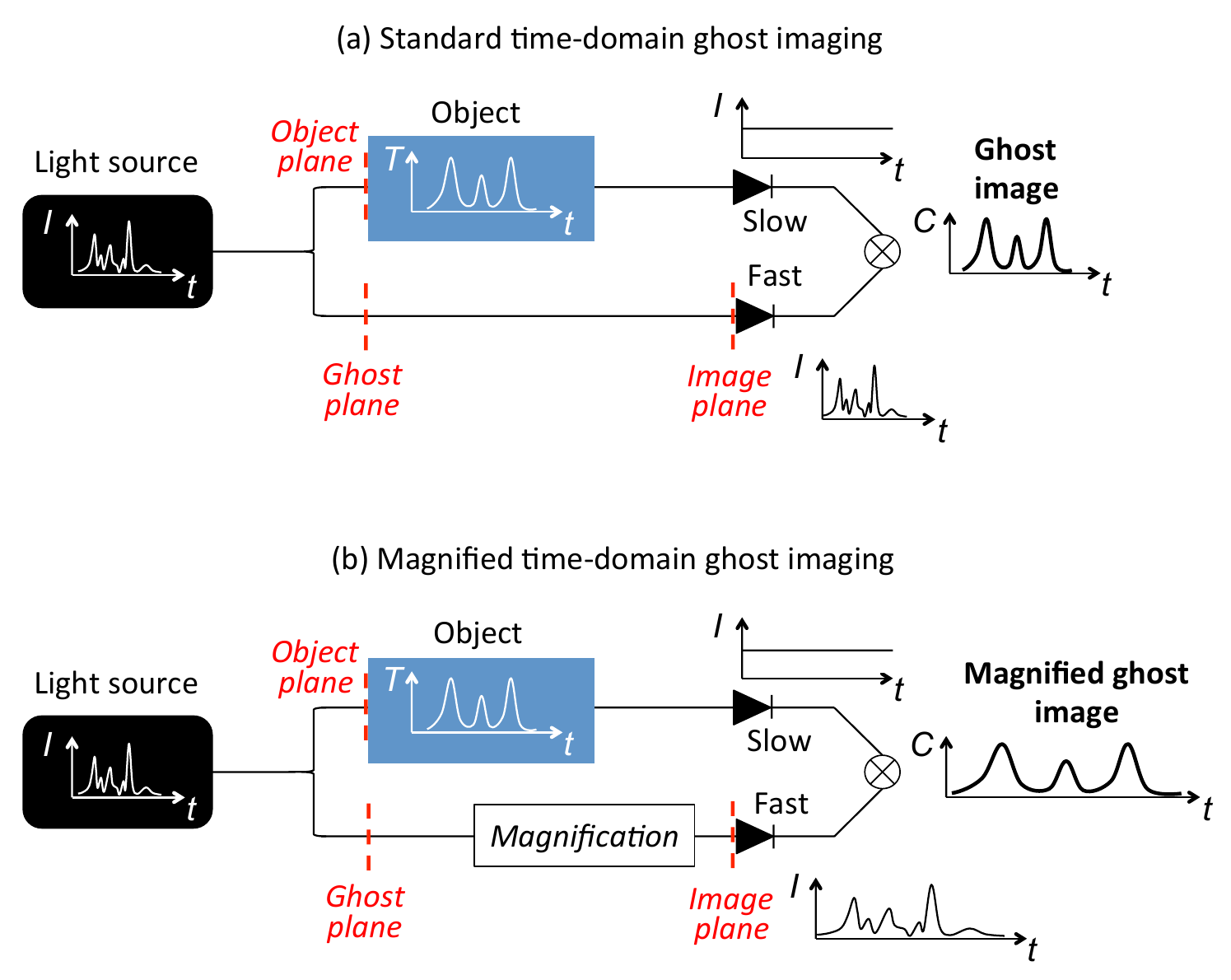}
	\caption{\label{fig:Operation_principle} Operation principle of (a) standard time-domain ghost imaging, and (b) magnified time- domain ghost imaging. The ghost plane is defined in the reference arm as the equivalent of the object plane (which is, by definition, located in the test arm), such that the dispersion accumulated by the light between the source and the ghost plane is equal to the dispersion accumulated between the source and the temporal object. {$I$~=~light} intensity. {$T$~=~transmission} of an intensity modulator~(=~object). {$C$~=~correlation} function. }
\end{figure}

In time-domain ghost imaging, the fast temporal fluctuations of an incoherent light source are divided between a test arm where a temporal object modulates the intensity fluctuations of the source, and a reference arm where the fluctuations are resolved in real time in the image plane\cite{Ryczkowski16} (i.e. the plane of the detector, see Fig. \ref{fig:Operation_principle}). By correlating the time-resolved fluctuations from the reference arm with the total (integrated) power transmitted in the test arm, a perfect copy of the temporal object can be retrieved (see Fig. \ref{fig:Operation_principle}(a)).
The temporally incoherent light source may be a quasi-continuous wave source with a fluctuation time inversely proportional to the source bandwidth, or a pulsed source with large intensity variations within a single pulse and from pulse to pulse\cite{Shirai10}. The correlation is calculated from multiple measurements synchronized with the temporal object. Note that the average intensity profile of the source over the measurement time window does not affect the ghost image. However, if the magnitude of the source intensity fluctuations varies over the duration of the temporal object (which can be the case especially for a pulsed source), the ghost image is distorted and requires post-processing correction\cite{Shirai10}.

In order to obtain a magnified ghost image, the temporal fluctuations of the source in the reference arm must be magnified\cite{Shirai10} whilst in the test arm one only needs to measure the total (integrated) intensity with no modification compared to standard time-domain ghost imaging (see Fig. \ref{fig:Operation_principle}(b). Magnification of the source fluctuations can be, in principle, obtained using a time lens system\cite{Salem08,Foster08,Schroder10}. However, time lens systems generally require complicated schemes to impose the necessary quadratic chirp onto the signal to be magnified and typically operate only at a fixed repetition rate with limited numerical apertures.

\begin{figure*}
	\centering
	\includegraphics[width=.7\textwidth]{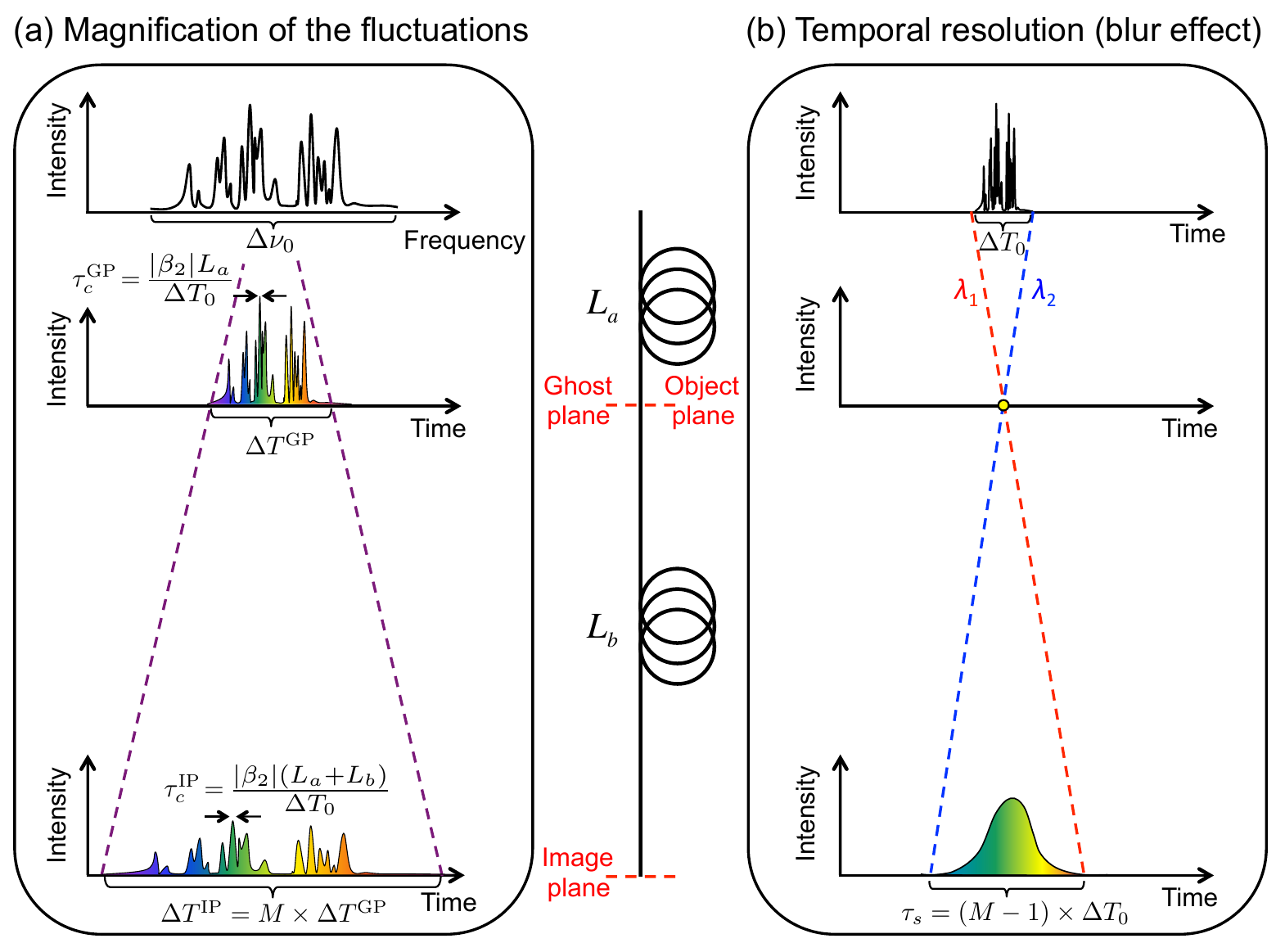}
	\caption{Spectrum-to-time transformation of the incoherent supercontinuum (SC). (a) Temporal magnification of the intensity fluctuations of the SC. (b) Temporal resolution limit from the finite duration of the SC pulses. $\Delta T_0$ and $\Delta \nu_0$ represents the initial duration and bandwidth of the SC pulses, respectively. $\Delta T^{GP}$, and $\Delta T^{IP}$ represent the duration of the SC pulses at the ghost and image planes, respectively. $\tau_c^{GP}$ and $\tau_c^{IP}$ denote the characteristic fluctuation time within each SC pulse at the ghost and image planes, respectively. $\tau_s$ is the temporal blur resulting from the different spectral components $\lambda_1$ and $\lambda_2$ that corresponds to the temporal edges of the initial SC pulses and temporally overlap in the ghost plane.}
	\label{fig:geometric_scheme}
\end{figure*}

A more straightforward approach consists in using spectrum-to-time transformation of the random shot-to-shot spectral fluctuations of a pulsed incoherent light source as illustrated in Fig. \ref{fig:geometric_scheme}. Because the SC is incoherent, the characteristic frequency of the spectral fluctuations is well-approximated by $\Delta\omega_c \approx 1 / \Delta T_0 $ where $\Delta T_0$ is the average duration of the SC pulse. These spectral fluctuations are first converted into the time domain using a dispersive fiber with total dispersion $\beta_2L_a$, resulting in pulses with (intra-pulse) temporal fluctuations $\tau_c^{\text{GP}}\approx |\beta_2|L_a / \Delta T_0$ at the ghost plane (defined as the equivalent of the object plane in the reference arm, see Fig. \ref{fig:Operation_principle}). These fluctuations are then divided between the test arm where the temporal object is located and the reference arm where they are stretched further in another dispersive fiber with total dispersion $\beta_2L_b$. The fluctuation time at the image plane (i.e. after propagation in the second dispersive fiber of length $L_B$) is $\tau_c^{IP} \approx |\beta_2|(L_a+L_b) / \Delta T_0$ , such that the temporal fluctuations in the image plane are magnified by a factor $M = (\beta_2L_a +\beta_2L_b)/\beta_2L_a = 1+L_b/L_a$ compared to the fluctuations in the ghost plane (see Fig. \ref{fig:geometric_scheme}(a). 

By correlating the magnified random fluctuations measured in the reference arm with the total transmitted intensity through the object in the test arm, one then directly obtains an \textit{M}-time magnified image of the temporal object. 
%
The initial duration $\Delta T_0$ of the SC pulses is finite such that each time instant in the ghost plane (and, equivalently in the test arm, each time instant of the temporal object) actually includes the contribution from several spectral components. These spectral components propagate to the image plane with different group-velocities due to dispersion (see Fig. \ref{fig:geometric_scheme}(b), which results in a "temporal blur effect" that limits the resolution of the imaging system. The temporal blur is defined as the delay $\tau_s$, in the image plane, between the frequencies corresponding to the temporal edges of the initial SC pulses and contributing to the same time instant in the ghost plane (see Supplement 1). Basic geometric considerations in Fig. \ref{fig:geometric_scheme}(b) show that:
\begin{equation}
\tau_s = \frac{L_b}{L_a} \Delta T_0 = (M-1)\Delta T_0
\end{equation}

For each SC pulse $i$, the oscilloscope records a pair of measurements: the magnified fluctuations $I_{\text{ref}}^{(i)}(t)$, and the total intensity transmitted through the electro-optic modulator $I_{\text{test}}^{(i)}$. This pair is recorded $N$ times, and the normalized correlation function which produces the ghost image is then calculated according to:
\begin{equation}
C(t) = \frac
{\left< \Delta I_{\text{ref}}^{(i)}(t) \cdot \Delta I_{\text{test}}^{(i)} \right>}
{\sqrt{\left< \bigg [\Delta I_{\text{ref}}^{(i)}(t)\bigg ]^2 \right> \left< \bigg[\Delta I_{\text{test}}^{(i)}\bigg]^2 \right>}}
\end{equation}
where $\left < \right>$ represents the ensemble average over the $N$ realizations ($i = 1\dots N$), and $\Delta I^{(i)}= I^{(i)}- \big <I^{(i)}\big >$.

\section{Experimental setup}
\begin{figure*}
	\centering
	\includegraphics[width = .7\linewidth]{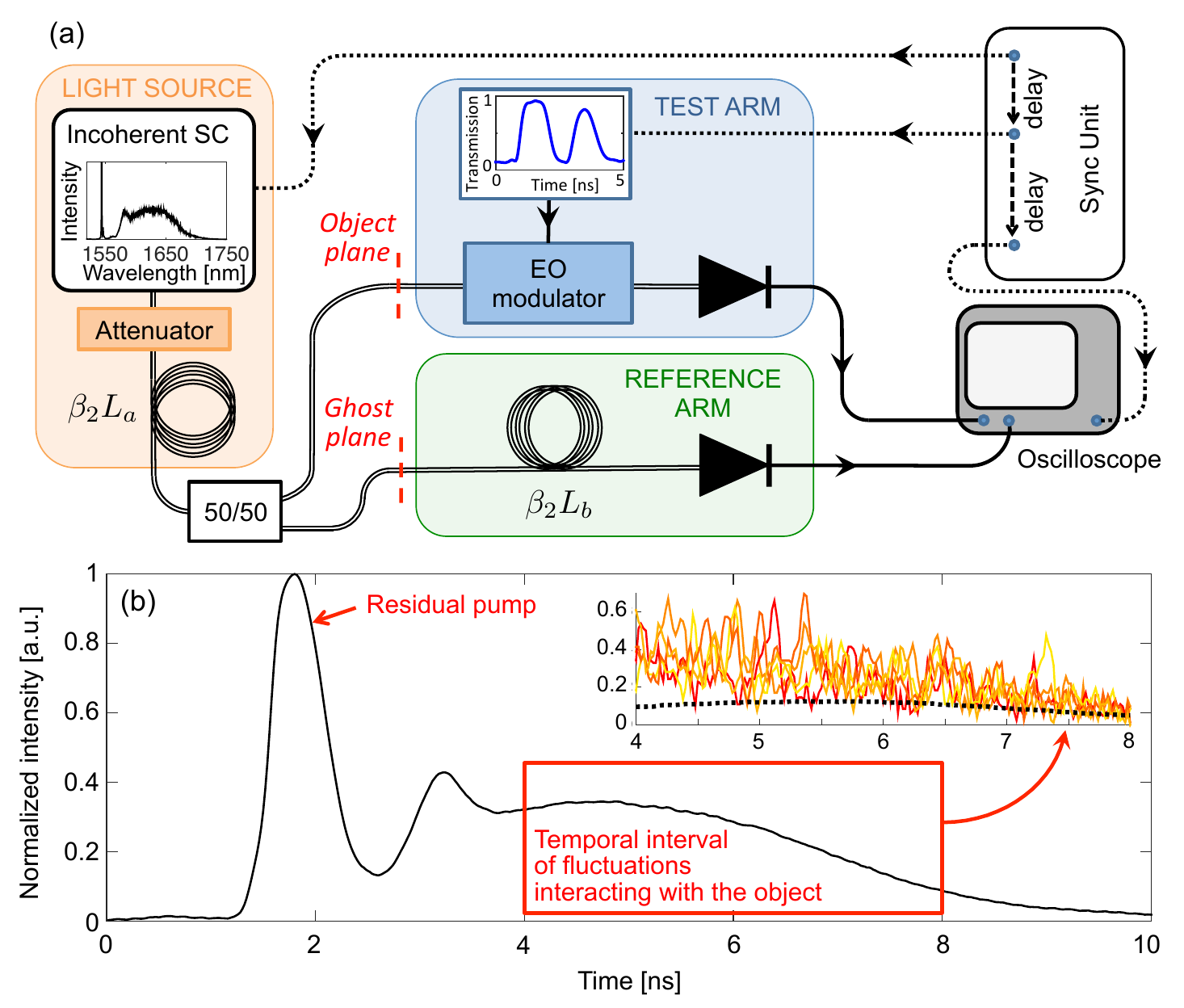}
	\caption{Magnified time-domain ghost imaging experimental setup. (a) Setup. The synchronization unit allows to synchronize the electric signal driving the transmission function of the electro-optic (EO) modulator with the arrival time of the SC pulse at the modulator and trigger the oscilloscope. The inset in the “Light source” box shows the average SC spectrum as measured with an optical spectrum analyzer with 0.1-nm resolution. (b) Average of 10 000 temporal traces recorded (with a 12.5-GHz InGaAs photodiode, Electro-Optics Technology ET-3500F) at the ghost plane, showing that, after propagating in the 2.5-km fiber, the temporal profile of the SC reproduces its spectral shape (spectrum-to-time transformation). Inset: Examples of 5 distinct SC pulse traces recorded at the ghost plane together with the standard deviation (dotted black line) calculated over 10 000 realizations.}
	\label{fig:experimental_setup}
\end{figure*}

The experimental setup is illustrated in Fig. \ref{fig:experimental_setup}(a). The light source is a spectrally filtered incoherent SC with large shot-to-shot spectral fluctuations. It is generated by injecting 0.5-ns pulses produced by an Erbium-doped fiber laser (Keopsys PEFL-KULT) operating at 1547 nm with 100-kHz repetition rate into the anomalous dispersion regime of a 6-m long dispersion-shifted fiber (Corning ITU-T G.655) with zero-dispersion at 1510 nm. The spectral components of the resulting SC below 1550 nm are filtered out with a long-pass filter to obtain a relatively flat spectrum. The average power of the SC is then reduced with an attenuator (Thorlabs VOA50-FC) to avoid any nonlinear processes that may occur during further propagation in an optical fiber.

The spectral fluctuations are converted into the time domain using an SMF-28 fiber of length $L_a = 2.5 ~\text{km}$ and dispersion parameter $\beta_2 = -20~ \text{ps}^2\text{/km at } 1550 ~\text{nm}$. They are then split between the test and reference arm with a 50/50 coupler. In the test arm, the temporal object is the transmission of a zero-chirp 10-GHz-bandwidth electro-optic modulator (Thorlabs \mbox{LN81S-FC}) driven by a programmable nanosecond pulse generator (iC-Haus iC149). It consists of two {0.75-ns} pulses with different amplitudes, spanning a total duration of {3.5~ns}. In the reference arm, the temporal fluctuations are magnified with an additional SMF-28 fiber of length $L_b=10~\text{km}$.

The detector in the test arm is a 5-GHz InGaAs photodiode (Thorlabs DET08CFC/M) whose response is integrated over 5~ns, such that the effective bandwidth is equal to 0.2 GHz only and the temporal profile of the object cannot be resolved in the test arm. The detector in the reference arm is a 1.2-GHz InGaAs photodiode (Thorlabs DET01CFC). The intensities measured by the two detectors are recorded by a real-time oscilloscope (Tektronix DSA72004). The detection bandwidth was intentionally limited to 625~MHz (with a sampling rate of 6.25~GS/s). Thus, the effective response time of the detection system that measures the fluctuations in real time in the reference arm is $\tau_d$ = 1.6~ns.

\section{Results and Discussion}
Initially (i.e. immediately after the spectral filtering stage), the SC has a bandwidth of 80~nm and the average duration of the SC pulses $\Delta T_0$  was measured to be less than 200~ps. Note that the duration of the SC after filtering is shorter than the original pump pulses. 
At the ghost and object planes (i.e. after the first {2.5-km} dispersive stage), the average duration of the SC pulses $\Delta T^{\text{GP}}$ was measured to be c.a. 4~ns. The standard deviation of the magnitude of the fluctuations is nearly constant over this time span (see the dotted black curve in the inset of Fig.  \ref{fig:experimental_setup}(b), such that the ghost image will not be distorted. 
At the image plane (i.e. after the second 10-km dispersive stage), the average duration of the SC pulses $\Delta T^{\text{IP}}$  was measured to be c.a. 20~ns, a 5 times increase compared to the original duration, as expected.

\begin{figure}
	\includegraphics[width=.7\linewidth]{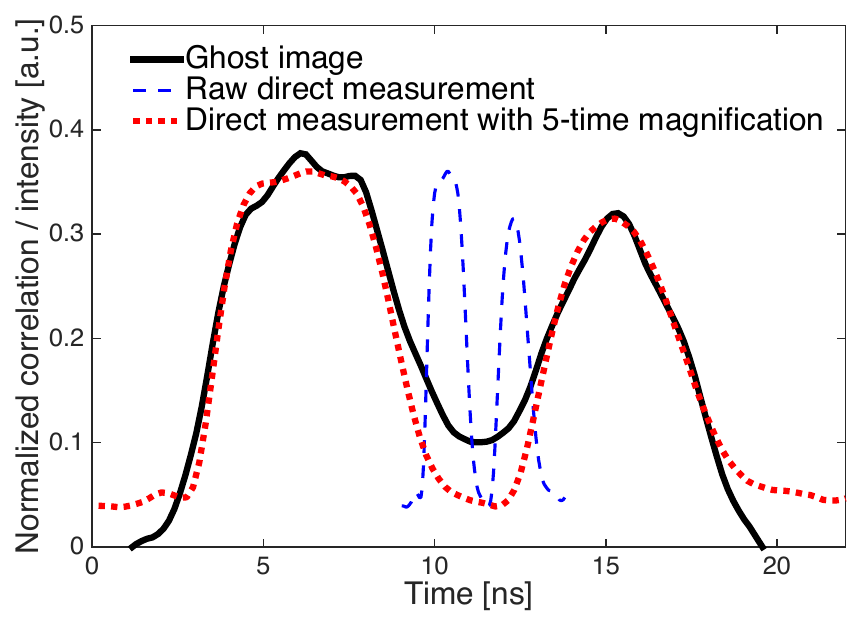}
	\caption{Magnified ghost image obtained by correlating the signals of the test and reference arms over 100~000 supercontinuum pulses (solid black line). For comparison, the direct measurement of the temporal object (with a 5-GHz detector) is shown before (dashed blue line) and after (dotted red line) post- processed 5-time magnification.}
	\label{fig:ghost_image}
\end{figure}

The correlation {$C(t)$} calculated over $N = 100~000$ SC pulses allows us to construct a ghost image magnified by a factor of $M= 5$, as shown in Fig. \ref{fig:ghost_image}. In this figure, we compare the ghost image with the original temporal object measured directly with a continuous-wave laser and 5-GHz photodiode (Thorlabs DET08CFC/M) and magnified 5 times through post-processing. We can see excellent agreement, both in terms of duration and amplitudes ratio of the object pulses, confirming the 5-time magnification factor of the object duration in the ghost imaging configuration.

\begin{figure}[ht]
	\includegraphics[width=.7\linewidth]{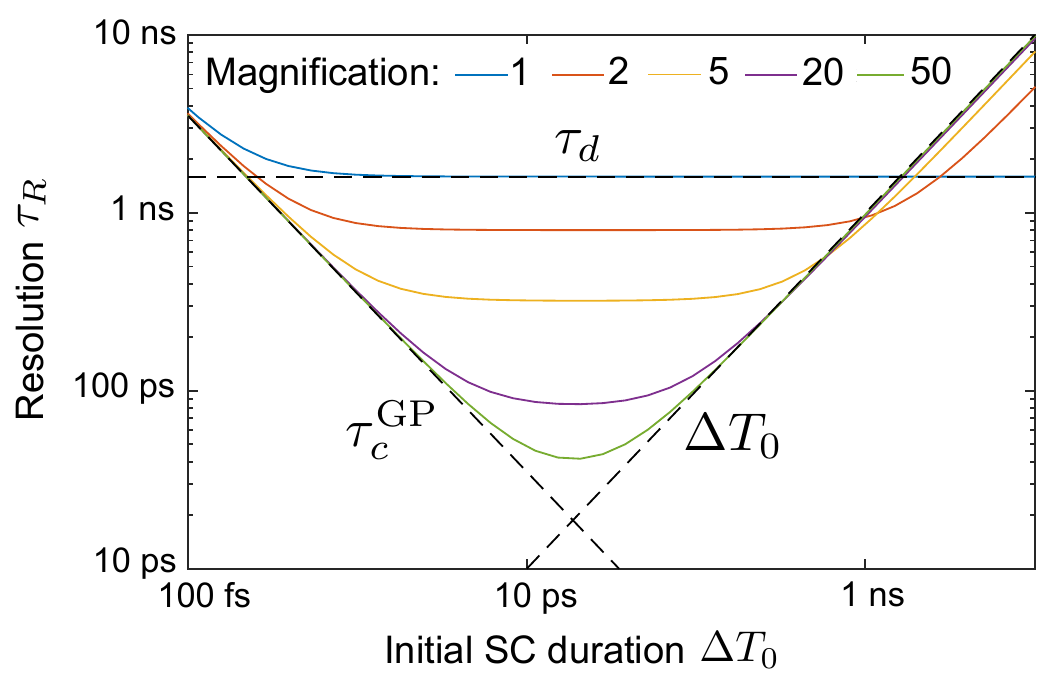}
	\caption{Resolution of the magnified ghost imaging setup as a function of the supercontinuum initial duration and for various magnification factors. The time response of the detection system in the reference arm $\tau_d$ is taken to be 1.6~ns, and the characteristic fluctuation time of the supercontinuum pulses in the ghost plane $\tau_c^{\text{GP}}$ is set equal to $50~ ps^2/\Delta T_0$ (consistent with our experimental parameters). The dashed lines illustrates the different factors that limit the resolution (detector speed $\tau_d$, characteristic time of fluctuations at the ghost plane $\tau_c^{\text{GP}}$ and initial duration of the SC pulses $\Delta T_0$.}
	\label{fig:resolution}
\end{figure}

There are some constraints which need to be considered for optimum resolution of the temporal object. Firstly, the time span of the fluctuations in the ghost plane (i.e. the duration of the SC pulses after the first dispersive fiber $\Delta T^{\text{GP}}$) needs to be longer or equal to that of the temporal object to be retrieved. This criterion is actually satisfied in our experiment, since $\Delta T^{\text{GP}}\approx$ 4 ns and the total duration of the object is only 3.5~ns. Secondly, the characteristic fluctuation time within each SC pulse at the ghost plane $\tau_c^{\text{GP}}$ needs to be shorter than the shortest object detail that one wishes to resolve. The temporal resolution $\tau_R$ of the imaging scheme is then determined by the combination of (i) the time response $\tau_d$ of the detection system, (ii) the characteristic time $\tau_c^{\text{GP}}$  of the random intensity fluctuations in each SC pulse at the ghost plane, and (iii) the initial duration$\Delta T_0$ of each SC pulse (i.e. before the spectrum-to-time transformation) which induces a temporal blur $\tau_s =(M-1)\Delta T_0$ in the image plane, as discussed in the previous section (see Fig. \ref{fig:geometric_scheme}(b). The overall resolution can then be approximated as
\begin{equation}
\tau_R = \sqrt{\left( \frac{\tau_d}{M} \right )^2 +\left(\tau_c^{\text{GP}} \right )^2+\left(\frac{\tau_s}{M} \right )^2}
\end{equation}
The resolution of the magnified ghost imaging system is illustrated in Fig. \ref{fig:resolution} as a function of the initial SC pulse duration $\Delta T_0$ and for different values of the magnification factor. We can see that, for short initial durations {($\leq$~1~ps)}, it is the fluctuation time at the ghost plane $\tau_c^{\text{GP}}$  that determines the overall resolution of the imaging system. In contrast, for long SC pulse durations {($\geq$~100~ps)}, it is the time delay $\tau_s$ between the SC frequencies at the image plane that sets the temporal resolution. The response time of the detection system $\tau_d$ only has an effect for small magnification factors. The temporal resolution $\tau_R$ in the results of Fig. \ref{fig:ghost_image} is estimated to be 360 ps, determined both by the resolution of the detection system in the reference arm $\tau_d = 1.6 ~ns$ and by the temporal spreading of the SC frequencies in the image plane $\tau_s\approx 0.8 ~ \text{ns}$, the fluctuation time of the SC pulses at the ghost plane $\tau_c^{\text{GP}}\approx  0.3 ~\text{ps}$ having a negligible influence. The resolution of the imaging system is therefore improved by a factor $\tau_d/\tau_R$ approximately equal to the magnification factor $M$ compared to the standard ghost imaging setup.

\section{Conclusions}
Using dispersive spectrum-to-time transformation of the fluctuations of an incoherent supercontinuum we have demonstrated ghost imaging with magnification in the time domain. This approach can pave the way for overcoming the limited resolution of the standard time-domain ghost imaging whilst requiring only simple modifications of the experimental setup. We emphasize that the magnified approach demonstrated here is also insensitive to any distortion that would affect the light field after the object. Our results open novel perspectives for dynamic imaging of ultrafast waveforms with potential applications in communications and spectroscopy.
\section*{Funding Information}
G. G. and A. T. F gratefully acknowledge support from the Academy of Finland (project 132279). J. M. D. acknowledges the ERC project MULTIWAVE.


%
%

%



\bibliography{references}

\end{document}